\documentstyle[12pt]{article}

\parskip=\medskipamount % space between paragraphs (LaTeX)
\renewcommand {\c}  {\'{c}}
\newcommand {\cc} {\v{c}}

\newcommand{\beq}{\begin{equation}}
\newcommand{\eeq}{\end{equation}}
\newcommand{\bdm}{\begin{displaymath}}
\newcommand{\edm}{\end{displaymath}}
\newcommand{\beqa}{\begin{eqnarray}}
\newcommand{\eeqa}{\end{eqnarray}}
\newcommand{\beqab}{\begin{eqnarray*}}
\newcommand{\eeqab}{\end{eqnarray*}}
\def\nn{\nonumber}

 \def\@makefnmark{\hbox to 0pt{$^{\@thefnmark}$\hss}}  %ORIGINAL

\begin{document}

\renewcommand{\thesubsection}{\arabic{subsection}.}
\renewcommand{\theequation}{\arabic{equation}}

%***********************TITLE PAGE**************************************
%
%

\begin{titlepage}
\begin{center}
{\Large \bf  MINIMAL UNITARY MODELS AND THE CLOSED ${\bf SU(2)_{q}}$
INVARIANT SPIN CHAIN}\\
\vspace{3.0cm}
{\normalsize \bf Silvio Pallua}\footnote{e-mail: pallua@phy.hr}
and {\normalsize \bf  Predrag Prester}\footnote{e-mail: pprester@phy.hr}\\[5mm]
{\it Department of Theoretical Physics\\University of Zagreb, POB
162\\Bijeni\cc ka c.32, 41001 Zagreb, Croatia}\\
\vspace{3.0cm}
{\large \bf Abstract}
\end{center}
\vspace{0.5cm}
We consider the Hamiltonian of the closed $SU(2)_{q}$ invariant chain. We
project a particular
class of statistical models belonging to the unitary minimal series. A
particular model
corresponds to a particular value of the coupling constant. The operator
content is derived.
This class of models has charge-dependent boundary conditions. In simple cases
(Ising, 3-state
Potts) corresponding Hamiltonians are constructed. These are non-local as the
original spin
chain.
\end{titlepage}
%
%*****************************TEXT***************************************
\newpage
\subsection{Introduction}
\setcounter{equation}{0}

Quantum groups together with the Temperly-Lieb algebra play a particular
role in integrable spin chains [1]. On the other hand, it may be
interesting to study particular Hamiltonians which are invariant to
the quantum group [1,2,3,4,5]. The quantum group invariant
Hamiltonian for the closed spin chain was constructed by Martin
[6]. This model was independently investigated in [7] and [8]. It was
shown that the properties of the ground state were such that for special
values of the coupling constant, conformal anomalies of minimal unitary
theories were obtained. In addition, this Hamiltonian implied boundary
conditions which depended on the coupling constant (or quantum group
parameter $q$) and quantum numbers of the sector. This second property
made this Hamiltonian different from the XXZ chain with the toroidal
boundary condition where the twist was common to all sectors of a given
Hamiltonian [9,10,11]. In this paper we want to proceed with this
investigation and show that it is possible to project from the closed
quantum chain partition functions of  statistical models corresponding
to  minimal unitary theories. In the finite-size scaling limit, we obtain
the spectra and the operator content of these theories. For finite chains,
the spectra of these models can be related to the starting quantum chain. Like
the original XXZ chain, the projected systems also have sector-dependent
boundary conditions. In our derivation we shall try to exploit the
theory of representations of quantum groups [1,13,14] and the division
of all states in ``good'' and ``bad''. Keeping only ``good'' states will lead
us to unitary theories.

\subsection{Statistical systems and the quantum chain}

We start with the Hamiltonian for the closed $SU(2)_{q}$ invariant chain [5,7]
\beqa
H&=&Lq-\sum_{i=1}^{L-1} R_{i}-R_{0} \\
R_{0} &=& G R_{L-1} G^{-1} \\
G &=& R_{1}\cdots R_{L-1}
\eeqa
where $R_{i}$ are $4\times 4$ matrices
\beq
R_{i}=\sigma_{i}^{+}\sigma_{i+1}^{-}+\sigma_{i}^{-}\sigma_{i+1}^{+}+
\frac{q+q^{-1}}{4} (\sigma_{i}^{3}\sigma_{i+1}^{3}+1)-
\frac{q-q^{-1}}{4} (\sigma_{i}^{3}-\sigma_{i+1}^{3}-2)\;\; .
\eeq
We choose the quantum group parameter $q$ to be on the unit circle
\beqa
q&=&e^{i\varphi} \\ \frac{q+q^{-1}}{2}&=&\cos \varphi=-\cos \gamma\;\; .
\nonumber
\eeqa
The Hamiltonian is invariant to generators of the quantum group
\beqa
S^{3}&=&\frac{1}{2}\sum_{i=1}^{L}\sigma_{i}^{3} \\
S^{\pm}&=&\sum_{i=1}^{L} q^{-\sigma^{3}/2}\otimes
\cdots \otimes q^{-\sigma^{3}/2}\otimes\sigma_{i}^{\pm}\otimes
q^{\sigma^{3}/2}\otimes\cdots\otimes q^{\sigma^{3}/2}\;\; . \nonumber
\eeqa
The operator $G$ plays the role of the translation operator
\beq
G R_{i} G^{-1}=R_{i+1} \hspace{10 mm}R_{L}=R_{0}
\hspace{10 mm} i=1,\ldots ,L-1
\eeq
and also commutes with the Hamiltonian. We shall be interested in
the cases in which the quantum group parameter is a root of unity:
\beq
q^{n}=\pm 1 \;\; .
\eeq

We shall first study the generic irrational case. In this case, one
can decompose the space of states into the direct sum of irreducible
representations of the quantum group which are in one-to-one correspondence
with the usual $SU(2)$ representations. It is therefore sufficient
to treat the highest weight states. All other states can be obtained
with the action of the $S^{-}$ operator. We derived the Bethe
Ansatz (BA) equation in [7]. In this reference the energy eigenvalues are given
by
\beq
E=2\sum_{i=1}^{M}(\cos\varphi -\cos k_{i})\hspace{10 mm} M=\frac{L}{2}-Q \;\; .
\eeq
Here $Q$ is the eigenvalue of $S^{3}$ and $k_{i}$ satisfy the BA constraints
\beq
Lk_{i}=2\pi I_{i}+2\varphi (Q+1)-\sum_{\stackrel{\scriptstyle j=1}{j\neq
i}}^{M}\Theta (k_{i},k_{j})\; \hspace{10 mm} k_{i}\neq \varphi
\eeq
where $I_{i}$ are integers (half-integers) if $M$ is odd (even), and $\Theta
(k_{i},k_{j})$ is the usual two-particle phase defined in [7].

It is important to notice that the BA functions $\Psi_{M} (n_{1},\ldots,n_{M})$
satisfy non-trivial boundary conditions:
\beq
\Psi_{M} (n_{2},\ldots,n_{M},n_{1}+L)=e^{\imath \phi}\Psi_{M}
(n_{1},\ldots,n_{M})
\eeq
where the numbers $n_{i}$ denote the positions of down spins and
\beq
\phi=2\varphi(Q+1)\;\; .
\eeq
This means that quantum invariance implies a non-trivial boundary
condition. This boundary condition has two properties. It depends on
the coupling constant
\beq
\gamma=\pi-\varphi
\eeq
and on the sector defined by the charge $Q$.

Owing to the antisymmetry of phase shifts, from (10) it follows that
\beq
\sum_{i=1}^{M} k_{i}=\frac{2\pi}{L}\sum_{i=1}^{M} I_{i}+\frac{2M}{L}
\varphi(Q+1)\;\; .
\eeq
This allows us to determine the eigenvalues of the translation operator
$G$ or equivalently of the operator $P$
\beq
P=\imath \ln G \;\; .
\eeq
In fact,
\beqa
P&=&\sum_{i=1}^{M} k_{i}-\varphi\left(Q-1+\frac{L}{2}\right)
\nonumber\\&=&\frac{2\pi}{L}
\sum_{i=1}^{M}I_{i}+\varphi\left[-\frac{L}{2}-Q+1+\frac{2M}{L}(Q+1)\right] \;\;
{}.
\eeqa

It was also shown in [7] that the finite-size correction to the
thermodynamic limit of the ground-state energy was given by ($L$ even)
\beq
E_{0}(L)=E_{0}(\infty)-\frac{\pi c\,\zeta}{6L}+O\left(\frac{1}{L}\right)
\eeq
where
\beq
\zeta=\frac{\pi\sin\gamma}{\gamma}\;\; .
\eeq
The conformal anomaly $c$ was found to be
\beq
c=1-\frac{6(\pi-\varphi)^{2}}{\pi\varphi}
\eeq
for $\varphi\in [\frac{\pi}{2},\pi]$. We are particularly interested
in the values
\beq
\varphi=\frac{\pi m}{m+1}\hspace{10 mm} m=3,4,\ldots \;\; .
\eeq
because they give the conformal anomalies of the minimal unitary models:
\beq
c=1-\frac{6}{m(m+1)}\hspace{10 mm} m=3,4,\ldots
\eeq
Now we define scaled gaps
\beqa
\overline{E}_{n}&=&\frac{L}{2\pi\zeta}(E_{n}-E_{0}) \\
\overline{P}_{n}&=&\frac{L}{2\pi}(P_{n}-P_{0}+\varphi\, Q)\;\; .
\eeqa
We introduce the partition function in some sector $Q\geq 0$ :
\beq
{\cal
F}_{Q}(z,\overline{z},L)=\sum_{all\;states}z^{\frac{1}{2}(\overline{E}_{n}
+\overline{P}_{n})}\,\overline{z}^{\frac{1}{2}(\overline{E}_{n}-\overline{P}_{n})}
\eeq
One can also introduce the partition function ${\cal K}_{Q}(z,\overline{z},L)$
for
the highest-weight states in the generic case with
\beq
{\cal K}_{Q}(z,\overline{z},L)=\sum_{\stackrel{\scriptstyle highest}{weight\;
states}}
z^{\frac{1}{2}(\overline{E}_{n}+\overline{P}_{n})}\,\overline{z}^{\frac{1}{2}
(\overline{E}_{n}-\overline{P}_{n})}\;\; .
\eeq
The function ${\cal K}_{Q}$ can also be expresed as
\beq
{\cal K}_{Q}(z,\overline{z},L)={\cal F}_{Q}(z,\overline{z},L)-{\cal F}_{Q+1}(z,
\overline{z},L)
\hspace{10 mm} 0\leq Q\leq \frac{L}{2}\;\; .
\eeq
This relation can be inverted into
\beq
{\cal F}_{Q}(z,\overline{z},L)=\sum_{j=Q}^{\frac{L}{2}}{\cal
K}_{j}(z,\overline{z},L)\;\; .
\eeq
The partition function for the particular case (8), when $q$ is a
root of unity, can be obtained by continuity from the generic case.
It is known [1] that, in this case, some representations will mix in
higher dimensional representations (``bad'' representations) which will
contain subrepresentations of zero norm. There will however still
exist representations isomorphic to the usual $SU(2)$ representations
with a non-vanishing norm (``good'' representations). We can therefore
expect that the ``good'' sector will lead us to interesting physical
models. We therefore need an expression for the partition function
${\cal D}_{Q}(z,\overline{z},L)$ for the highest-weight states from the
``good'' sector\footnote{These states can also be characterised (relation
(1.19)
in [1]) by the condition that they belong to the kernel of $S^{+}$
and do not belong to the image of $(S^{+})^{n-1}$.}. This formula was
derived by Pasquier and Saleur (relation (2.9) in [1]) in the context
of the open quantum chain. However, their arguments are based purely on
group- theoretical grounds and can also be repeated here with the same
result. Thus,
\beqa
{\cal D}_{Q}(z,\overline{z},L)&=&\sum_{r\geq 0}\left({\cal
K}_{Q+nr}(z,\overline{z},L)-{\cal
K}_{n-1-Q+nr}(z,\overline{z},L)\right)\hspace{6 mm} 0\leq Q\leq\frac{1}{2}
(n-1) \nonumber \\
&=&\sum_{r\geq 0}{\cal K}_{Q+nr}(z,\overline{z},L)-\sum_{r>0}{\cal K}_{-Q-1+nr}
(z,\overline{z},L)
\eeqa
where from (8) and (20) it follows that $n=m+1$.
For later convenience, we transform this formula into another form.
We denote the generating function of lowest weight states by
${\cal K}_{j}$ for $j<0$. Owing to the symmetries of the Hamiltonian we have
\beq
{\cal K}_{j}(z,\overline{z},L)={\cal K}_{-j}(z,\overline{z},L)\;\; .
\eeq
Then (28) can be written as
\beq
{\cal D}_{Q}(z,\overline{z},L)=\sum_{r\geq 0}{\cal K}_{Q+nr}(z,\overline{z},L)-
\sum_{r<0}{\cal K}_{Q+1+nr}(z,\overline{z},L)\;\; .
\eeq
Analogously to (26), we can express ${\cal K}_{-|j|}(z,\overline{z},L)$ as
\beq
{\cal K}_{-|j|}(z,\overline{z},L)={\cal F}_{-|j|}(z,\overline{z},L)-{\cal
F}_{-|j|-1}(z,\overline{z},L)\;\; .
\eeq
Using (26) and (31) in (30) we obtain
\beqa
{\cal D}_{Q}(z,\overline{z},L)=\sum_{r\geq 0}\left({\cal
F}_{Q+nr}(z,\overline{z},L)-{\cal F}_{Q+1+nr}(z,\overline{z},L)\right)\nonumber
\\ -\sum_{r<0}\left(
{\cal F}_{Q+1+nr}(z,\overline{z},L)-{\cal
F}_{Q+nr}(z,\overline{z},L)\right)\;\; .
\eeqa
It is convenient to introduce the notation
\beq
{\cal G}_{Q}(z,\overline{z},L)=\sum_{r\in Z}{\cal
F}_{Q+nr}(z,\overline{z},L)\;\; .
\eeq
With this notation, the generating function for the ``good'' sector
can be written as
\beq
{\cal D}_{Q}(z,\overline{z},L)={\cal G}_{Q}(z,\overline{z},L)-{\cal G}_{Q+1}
(z,\overline{z},L)\;\; .
\eeq
We shall see that ${\cal D}_{Q}(z,\overline{z},L)$ will define the spectrum of
a
statistical model with non-trivial boundary conditions (sector-dependent).
The spectrum of this model is related to the spectrum of our starting
quantum chain with the help of (34). The same relation is true in the
finite-size
scaling limit. In this case, however, we shall be able to determine
explicit formulae for the operator content of the resulting model.

\subsection{Quantum chain and the XXZ chain with a toroidal
            boundary condition}

Boundary conditions of the quantum chain are sector dependent ((11),(12)).
One can raise the natural question how the spectrum of the quantum chain
is related to the chains with toroidal boundary conditions. As indicated
previously, we are particularly interested in the ``good'' part of the
spectrum of the quantum chain. It will turn out that the answer to the above
question will enable us to use the results of [10,11] on toroidal models.
They will provide us with necessary arguments to show from the relation (34)
the results anticipated at the end of the preceding section.

We remind the reader of the results for the toroidal case [10,11]. The
Hamiltonian
is defined by
\beq
H(q,\phi)=-\sum_{i=1}^{L}\left\{ \sigma_{i}^{+}\sigma_{i+1}^{-}+
\sigma_{i}^{-}\sigma_{i+1}^{+}+
\frac{q+q^{-1}}{4} (\sigma_{i}^{3}\sigma_{i+1}^{3})\right\}
\eeq
\beq
\frac{q+q^{-1}}{2}=\cos\varphi =-\cos\gamma
\eeq
and
\beq
\sigma_{L+1}^{\pm}=e^{\mp\imath\phi}\sigma_{1}^{\pm}\hspace{10 mm}
\phi\in (-\pi,\pi]\;\; .
\eeq
This Hamiltonian commutes with
\beq
S^{z}=\sum_{i=1}^{L}\sigma_{i}^{3}
\eeq
and with the translation operator
\beq
T=e^{-\imath\phi\sigma_{1}^{3}/2}P_{1}P_{2}\cdots P_{L-1}
\eeq
where $P_{i}$, $i=1,\ldots ,L-1$ are permutation operators
\beq
P_{i}=\sigma_{i}^{+}\sigma_{i+1}^{-}+\sigma_{i}^{-}\sigma_{i+1}^{+}+
\frac{1}{2}\left( \sigma_{i}^{3}\sigma_{i+1}^{3}+1\right)\;\; .
\eeq
The momentum operator is then
\beq
P=\imath\ln T \;\; .
\eeq
The BA constraints for this system are [9]
\beq
Lk_{i}=2\pi I_{i}+\phi -\sum_{\stackrel{\scriptstyle j=1}{j\neq
i}}^{M}\Theta (k_{i},k_{j})\hspace{10 mm} i=1,\ldots,M
\eeq
and give
\beqa
E&=&-\frac{L}{2}\cos\varphi+2\sum_{i=1}^{M}(\cos\varphi-\cos k_{i}) \\
P&=&\sum_{i=1}^{M}k_{i}=\frac{2\pi}{L}\sum_{i=1}^{M}I_{i}+\frac{M}{L}\phi\;\; .
\eeqa
We define
\beq
\phi=2\pi l\;,\hspace{10 mm} -\frac{1}{2}<l\leq \frac{1}{2} \;\; .
\eeq
The finite-size scaling limit of this system is described by the $c=1$
conformal field theory of the compactified free-boson system with the
compactification radius
\beq
R^{2}=8h
\eeq
where
\beq
h=\frac{1}{4(1-\gamma/\pi)}
\eeq
and $h\geq\frac{1}{4}$.

Let us denote by $E_{Q;j}^{l}(L)$ and $P_{Q;j}^{l}(L)$ the eigenvalues
of $H$ and $P$ in the sector $S^{z}=Q$ with a boundary condition defined
by $\phi=2\pi l$. Then, following references [10,11], we can write
the expression for the finite-size scaling function of $H_{Q}^{l}$ :
\beqa
{\cal E}_{Q}^{l}(z,\overline{z})&=&\lim_{L\rightarrow\infty}
{\cal E}_{Q}^{l}(z,\overline{z},L) \nonumber \\ &=&\lim_{L\rightarrow
\infty}\sum_{j=1}^{(\begin{array}{c} {\scriptscriptstyle L} \\[-4 mm]
{\scriptscriptstyle Q+L/2} \end{array}
)}z^{\frac{1}{2}(\overline{E}_{Q;j}^{l}(L)+\overline{P}_{Q;j}^{l}(L))}\,
\overline{z}^{\frac{1}{2}(\overline{E}_{Q;j}^{l}(L)-\overline{P}_{Q;j}^{l}(L))}
\\
&=&\sum_{\nu\in Z}z^{\frac{[Q+4h(l+\nu)]^{2}}{16h}}\,\overline{z}^{
\frac{[Q-4h(l+\nu)]^{2}}{16h}}\prod_{n=1}^{\infty}(1-z^{n})^{-1}
(1-\overline{z}^{n})^{-1} \;\; .\nonumber
\eeqa
The symbols $\overline{E}_{Q;j}^{l}(L)$ and $\overline{P}_{Q;j}^{l}(L)$
denote the scaled gaps
\beqab
\overline{E}_{Q;j}^{l}(L)&=&\frac{L}{2\pi}(E_{Q;j}^{l}(L)-E_{0;0}^{0}(L)) \\
\overline{P}_{Q;j}^{l}(L)&=&\frac{L}{2\pi}P_{Q;j}^{l}(L) \;\; .
\eeqab
It was shown [10,11] that it was possible to project theories with $c<1$ by
choosing a new ground state with energy $E_{0;j_{0}}^{l_{0}}(L)$. The
number $j_{0}\geq1$ was chosen in such a way that the new ground state gave
the contribution $(z\overline{z})^{h(l_{0}+\nu_{0})^{2}}$ in the partition
function (48). The quantity $(l_{0}+\nu_{0})$ is related to $h$ by the
condition
\beq
c=1-\frac{6}{m(m+1)}=1-24h(l_{0}+\nu_{0})^{2}
\eeq
where $-\frac{1}{2}<l_{0}\leq\frac{1}{2}$ and $\nu_{0}\in Z$. From
(49) it follows that
\beq
l_{0}+\nu_{0}=[4hm(m+1)]^{-\frac{1}{2}} \;\; .
\eeq
Now new scaled gaps can be defined as
\beqab
\overline{F}_{Q;j}^{k}(L)&=&\frac{L}{2\pi}\left( E_{Q;j}^{k(l_{0}+\nu_{0})}
(L)-E_{0;j_{0}}^{l_{0}}(L)\right) \\
\overline{P}_{Q;j}^{k}(L)&=&\frac{L}{2\pi}P_{Q;j}^{k(l_{0}+\nu_{0})}(L)\;\; .
\eeqab
The corresponding finite-size scaling partition function is
\beqa
{\cal F}_{Q}^{k}(z,\overline{z})&=&\lim_{L\rightarrow\infty}{\cal F}_{Q}^{k}(z,
\overline{z},L) \nonumber \\
&=&\lim_{L\rightarrow\infty}\sum_{j=1}^{(\begin{array}{c}
{\scriptscriptstyle L} \\[-4 mm] {\scriptscriptstyle Q+L/2} \end{array}
)}z^{\frac{1}{2}(\overline{F}_{Q;j}^{k}(L)+
\overline{P}_{Q;j}^{k}(L))}\,\overline{z}^{\frac{1}{2}(\overline{F}_{Q;j}^{k}(L)
-\overline{P}_{Q;j}^{k}(L))}\;\; .
\eeqa
The relation (49) gives $c$ as a function of two independent real
parameters, $h$ and $l_{0}+\nu_{0}$. According to [10], two classes of
$c<1$ models can be defined imposing the relation
\beq
l_{0}+\nu_{0}=\frac{1}{M}-\frac{M}{4h}\;\; .
\eeq
They are called $R$ models if $M>0$ ($R=M$) and $L$ models if $M<0$
($L=-M$). From (49) and (52) it follows that
\beqa
\varphi &=&\frac{\pi m}{R^{2}(m+1)} \hspace{25 mm} \mbox{$R$ models} \\
\varphi &=&\frac{\pi (m+1)}{L^{2}m} \hspace{26 mm} \mbox{$L$ models}\;\; .
\eeqa
Our goal is to evaluate (34) extracted from the quantum chain. By
equating (19) and (21) we obtain the following equation:
\bdm
c=1-\frac{6(\pi -\varphi)^{2}}{\pi\varphi}=1-\frac{6}{m(m+1)}\;\; .
\edm
The only solution of this equation for which (47) holds is given by
\beq
\varphi=\frac{\pi m}{m+1}\;\; .
\eeq
Thus, for our purpose, it is sufficient to take $R=1$ models for which (53)
and (55) will coincide. In this case,
\beq
l_{0}+\nu_{0}=\frac{1}{m+1}
\eeq
and the function ${\cal F}_{Q}^{k}(z,\overline{z},L)$ has the periodicity
properties
\beq
{\cal F}_{Q}^{k}(z,\overline{z},L)={\cal F}_{Q}^{k\pm n}(z,\overline{z},L)
\eeq
where the integer $n$ is given by
\beq
n=m+1 \;\; .
\eeq
Consider the function ${\cal G}_{Q}^{k}(z,\overline{z},L)$
\beq
{\cal G}_{Q}^{k}(z,\overline{z},L)=\sum_{\nu\in Z}{\cal F}_{Q+\nu n}^{k}
(z,\overline{z},L)
\eeq
which satisfies
\beq
{\cal G}_{Q\pm n}^{k}(z,\overline{z},L)={\cal G}_{Q}^{k\pm n}(z,
\overline{z},L)={\cal G}_{Q}^{k}(z,\overline{z},L)={\cal G}_{n-Q}^{n-k}
(z,\overline{z},L) \;\; .
\eeq
{}From (48), (49) and (59) one obtains [10,11]
\beqa
{\cal D}_{Q}^{k}(z,\overline{z})&\equiv& {\cal G}_{Q}^{k}(z,\overline{z})
-{\cal G}_{k}^{Q}(z,\overline{z}) \nonumber \\ &=&\sum_{r=1}^{m-1}
\chi_{r,k-Q}(z)\,\chi_{r,k+Q}(\overline{z})
\eeqa
where
\beq
1\leq k\leq m\;,\hspace{5 mm} |Q|\leq\min \{k-1,m-k\} \;\; .
\eeq
The symbols $\chi_{r,s}$ denote the character functions of irreducible
representations of
the Virasoro algebra with highest weights $\Delta_{r,s}$ given by
\beq
\Delta_{r,s}=\frac{[(m+1)r-ms]^{2}-1}{4m(m+1)}\;\; .
\eeq
The functions ${\cal F}_{Q}^{k}$ are partition functions of the toroidal
chain with the boundary condition
\beq
\phi=2\pi\frac{k}{n}=2\pi\frac{k}{m+1}\hspace{10 mm}\mbox{in the sector
$S^{z}=Q$}\;\; .
\eeq
On the other hand, we have seen that the quantum chain has boundary conditions
given by
\beq
\phi=2\varphi (Q+1)=2\pi\frac{m}{m+1}(Q+1)\pmod{2\pi}\;\; .
\eeq
Comparing (64) with (65), one obtains
\beq
k=-(Q+1)\pmod{n} =m-Q \pmod{n}\;\; .
\eeq
The highest-weight states for the quantum chain in the sector of charge $Q$
satisfy the same BA equations as the toroidal Hamiltonian with the boundary
condition (65). As a consequence, the energy and momenta of the states are
simply related (as follows by comparing (9) with (43), and (16) with
(44)) by
\beqa
E&=&E({\rm toroidal})+\frac{L}{2}\cos\varphi \\
P&=&P({\rm toroidal})-\varphi\left(Q-1+\frac{L}{2}\right) \;\; .
\eeqa
Using (66) and (61) we obtain
\beq
{\cal D}_{Q}^{m-Q}(z,\overline{z})={\cal D}_{Q}^{-(Q+1)}(z,\overline{z})
={\cal G}_{Q}^{-(Q+1)}(z,\overline{z})-{\cal
G}^{Q}_{-(Q+1)}(z,\overline{z})\;\; .
\eeq
However, the relation (34) for the quantum chain, after using the symmetry
property
\beq
{\cal G}_{Q}(z,\overline{z})={\cal G}_{-Q}(z,\overline{z})
\eeq
has the same form as (69). Indeed, we know from [1] that formula (69)
for the toroidal Hamiltonian projects states which are in the kernel of $S^{+}$
and not in the image of $(S^{+})^{n-1}$. However, that was also the case with
the relation (34). In fact, the left sides of these two relations are both
``good'' highest-weight states with the same charge and the same boundary
condition and satisfy the same BA equations. Adding the usual assumption
that all ``good'' highest-weight states are given by BA states, we
conclude that the left sides of (34) and (69) are equal. In other words,
\beq
{\cal D}_{Q}(z,\overline{z},L)={\cal D}_{Q}^{m-Q}(z,\overline{z},L)\;\; .
\eeq

\subsection{Unitary minimal models and the quantum chain}

One result of the preceding section is the relation (71) which in combination
with (61) leads to
\beq
{\cal D}_{Q}(z,\overline{z})=\lim_{L\rightarrow\infty}{\cal D}_{Q}
(z,\overline{z},L)=\sum_{r=1}^{m-1}\chi_{r,m-2Q}(z)\,\chi_{r,m}(\overline{z})
\eeq
\beq
0\leq Q<\frac{m}{2} \;\; .
\eeq
The right-hand side of (72) has the form of partition functions of
physical systems [15,16] where $\chi_{r,s}$ are denoted character
functions of highest-weight representations of the Virasoro algebra. The
quantum parameter $\varphi$ determines $m$:
\beq
\varphi=\frac{\pi m}{m+1} \;\; .
\eeq
Accordingly, the construction (72) gives the partition function
of a system which consists of a ``good'' subset of states of the original
quantum chain; this system has the conformal anomaly
\beq
c=1-\frac{6}{m(m+1)}\hspace{10 mm} m=3,4,\ldots
\eeq
and the operator content can be read from (72) with the help of
formula (63). Owing to (75) this system belongs to the unitary series.

We present some simple examples.

\subsubsection{$m=3$}
{}From (73) it follows that $Q=0,1$.
\beqa
{\cal D}_{0}(z,\overline{z})&=&\sum_{r=1}^{2}\chi_{r,3}(z)\,\chi_{r,3}
(\overline{z})=\sum_{r=1}^{2}(\Delta_{r,3},\overline{\Delta}_{r,3}) \nonumber
\\
&=&(0,0)+(1/2,1/2) \\
{\cal D}_{1}(z,\overline{z})&=&\sum_{r=1}^{2}\chi_{r,1}(z)\,\chi_{r,3}
(\overline{z})=\sum_{r=1}^{2}(\Delta_{r,1},\overline{\Delta}_{r,3}) \nonumber
\\
&=&(1/2,0)+(0,1/2) \;\; .
\eeqa
We have used the usual notation
\beq
\chi_{r,s}(z)\,\chi_{p,t}(\overline{z})\equiv (\Delta_{r,s},
\overline{\Delta}_{p,t}) \;\; .
\eeq
These functions can be identified with the partition functions for given
sectors of the Ising chain. In fact, consider the Hamiltonian
\beq
H=-\frac{1}{2}\sum_{j=1}^{L/2}(\sigma_{j}^{3}+\sigma_{j}^{1}
\sigma_{j+1}^{1})
\eeq
with the boundary conditions
\bdm
\sigma_{\frac{L}{2}+1}^{1}=(-1)^{\tilde{q}}\sigma_{1}^{1}
\hspace{10 mm} \tilde{q}=0,1 \;\; .
\edm
This Hamiltonian commutes with the operator $\Sigma$
\beq
\Sigma=\sigma_{1}^{3}\cdots\sigma_{L/2}^{3} \;\; .
\eeq
We use $T_{q}^{\tilde{q}}$ to denote the partition function for the boundary
condition defined by $\tilde{q}$ and the eigenvalue $(-1)^{q}$ of $\Sigma$.
Their
conformal content was obtained in [17]. By comparison,
\beqa
{\cal D}_{0}&=&T_{0}^{0}\nonumber \\ {\cal D}_{1}&=&T_{1}^{1} \;\; .
\eeqa
Thus we see that, e.g., $T_{0}^{1}$ and $T_{1}^{0}$ are not contained in
this construction. It would be interesting to construct a Hamiltonian
containing just these sectors. In fact, this is a non-local Hamiltonian
already discussed in [7,11] which we mention here for completeness:
\beq
H=-\frac{1}{2}\left\{ \sum_{j=1}^{\frac{L}{2}-1}(\sigma_{j}^{3}+
\sigma_{j}^{1}\sigma_{j+1}^{1})+\sigma_{L/2}^{3}+
\sigma_{L/2}^{1}\sigma_{1}^{1}\Sigma\right\} \;\; .
\eeq

\subsubsection{$m=4$}
Again, $Q=0,1$.
\beqa
{\cal D}_{0}&=&\sum_{r=1}^{3}(\Delta_{r,4},\overline{\Delta}_{r,4}) \nonumber
\\
&=&(0,0)+(7/16,7/16)+(3/2,3/2) \nonumber \\
{\cal D}_{1}&=&\sum_{r=1}^{3}(\Delta_{r,2},\overline{\Delta}_{r,4}) \nonumber
\\
&=&(3/5,0)+(3/80,7/16)+(1/10,3/2) \;\; .
\nonumber
\eeqa

\subsubsection{$m=5$}
$Q=0,1,2$
\beqa
{\cal D}_{0}&=&\sum_{r=1}^{4}(\Delta_{r,5},\overline{\Delta}_{r,5}) \nonumber
\\
&=&(0,0)+(2/5,2/5)+(7/5,7/5)+(3,3) \nn \\
{\cal D}_{1}&=&\sum_{r=1}^{4}(\Delta_{r,3},\overline{\Delta}_{r,5}) \nonumber
\\
&=&(1/15,2/5)+(2/3,0)+(1/15,7/5)+(2/3,3) \\
{\cal D}_{2}&=&\sum_{r=1}^{4}(\Delta_{r,1},\overline{\Delta}_{r,5}) \nonumber
\\
&=&(3,0)+(7/5,2/5)+(2/5,7/5)+(0,3) \;\; . \nonumber
\eeqa
These are partition functions of the 3-state Potts model whose Hamiltonian
is given by
\beq
H=-\frac{2}{3\sqrt{3}}\sum_{j=1}^{N}\left(\sigma_{j}+\sigma_{j}^{\dagger}+
\Gamma_{j}\Gamma_{j+1}^{\dagger}+\Gamma_{j}^{\dagger}\Gamma_{j+1}\right)
\eeq
\beq
\sigma=\left(\begin{array}{ccc} 1&0&0 \\ 0&\omega&0 \\ 0&0&\omega^{2}
\end{array}\right) \hspace{5 mm}\Gamma=\left(\begin{array}{ccc} 0&0&1 \\
1&0&0 \\ 0&1&0 \end{array}\right)\hspace{5 mm}\omega =e^{\frac{2\pi\imath
}{3}} \;\; .
\eeq
We introduce the partition functions $T_{q}^{\tilde{q}}$ corresponding to the
boundary condition
\beq
\Gamma_{N+1}=\omega^{\tilde{q}}\Gamma_{1}\hspace{10 mm}q=0,1,2
\eeq
and the sector $q$ of the charge operator
\beq
M=\sigma_{1}\cdots\sigma_{N} \;\; .
\eeq
Then, comparing (83) with the decomposition (2.37a) of [10], we obtain
\beqa
{\cal D}_{0}&=&T_{0,+}^{0} \nonumber \\
{\cal D}_{1}&=&T_{1}^{2}=T_{2}^{1} \\
{\cal D}_{2}&=&T_{0,-}^{0} \;\; . \nonumber
\eeqa
The functions ${\cal D}_{0}$ and ${\cal D}_{2}$  have the same boundary
condition
and the charge $Q$, and the notation $+,-$ distinguishes between two
one-dimensional
representations of $S_{3}$ (see, e.g. [10]). All other partition functions are
forbidden
in our model. In fact the Hamiltonian for this case can be constructed as
\beqa
H=-\frac{2}{3\sqrt{3}}\left\{ \sum_{j=1}^{N-1}\left(\sigma_{j}+\sigma_{j}^{
\dagger}+\Gamma_{j}\Gamma_{j+1}^{\dagger}+\Gamma_{j}^{\dagger}\Gamma_{j+1}
\right)\right. \nonumber \\ \left. +\sigma_{N}+\sigma_{N}^{\dagger}+\Gamma_{N}
\Gamma_{1}^{\dagger}M+
\Gamma_{N}^{\dagger}\Gamma_{1}M^{\dagger}\right\}
\eeqa
where $N=L/2$. As expected, it is again non-local and implies sector-dependent
boundary conditions. We note that by making the replacement $M\rightarrow
M^{\dagger}$ in (89), and after an obvious adjustment of multiplicative and
additive constants, we obtain the Hamiltonian from [7]. These two Hamiltonians
have the same energy spectrum, but the momenta of opposite sign and thus a
different
operator content.

Thus starting with the closed quantum chain, we have obtained
the finite-size scaling limit of the partition functions for definite
statistical systems. The corresponding operator content can be read from the
relation
(72) given the deformation parameter $q$ of the original quantum chain. For
finite chains, the explicit character formula is not available.
However, in that case, we are still in a position to relate the
spectra of the two theories through the relation (23):
\beq
{\cal D}_{Q}(z,\overline{z},L)={\cal G}_{Q}(z,\overline{z},L)-{\cal G}_{Q+1}
(z,\overline{z},L) \;\; .
\eeq
This relation for partition functions implies that the spectrum of our
statistical system is contained in the quantum chain and can be obtained from
(90). We remind that
\beq
{\cal G}_{Q}(z,\overline{z},L)=\sum_{r\in Z}{\cal F}_{Q+rn}(z,\overline{z},L)
\eeq
where ${\cal F}_{Q}(z,\overline{z},L)$ is the partition function of the
original system defined by (24).

For illustration, we present energies of the quantum chain with four sites
for $m=3$, $n=4$. The construction (90) then gives us energies for the
projected statistical system. This system was previously identified as the
Ising
chain with two sites and with the boundary conditions dependent on the sector
and defined in (81). The set of energies of the projected system is a
subset of the set of energies of the original system and are underlined in
table 1. These energies are indeed also energies of the Ising chain (82), as
can be checked numerically. Our numbers are a subset of numbers in table 3
in [10] for chains with toroidal boundary conditions. We have to expect this
since we have shown that the spectrum of the quantum chain is contained in the
union
of spectra of toroidal Hamiltonians.
\begin{table}
  \caption{Scaled energy gaps defined by (22) for the quantum chain
           with 4 sites, $m=3$, $n=4$. The levels which are underlined
           correspond to the Ising model (82) with 2 sites}
  \begin{tabular}{ccc} \hline
 \makebox[35 mm]{${\cal G}_{0}$}& \makebox[35 mm]{${\cal G}_{1}$}&
 \makebox[35 mm]{${\cal G}_{2}$} \\ \hline
 \underline{0.000000}& -                   & -              \\
    0.450158         & \underline{0.450158}& -              \\
    0.450158         & \underline{0.450158}& -              \\
 \underline{0.900316}& -                   & -              \\
    1.086778         & 1.086778            & 1.086778       \\
    1.086778         & 1.086778            & 1.086778       \\ \hline
  \end{tabular}
\end{table}

In table 2 we have illustrated some features for the $m=5$, $n=6$ case of
the 3-state Potts model with two sites and sector-dependent boundary
conditions.We remark that in both cases the allowed boundary conditions are
those permitted
by the symmetry on duality transformations [18].
\bdm
H_{q}^{\tilde{q}}=H_{\tilde{q}}^{q}
\edm
We shall consider this point elsewhere.
\begin{table}
  \caption{Scaled energy gaps defined by (22) for the quantum chain
           with 4 sites, $m=5$, $n=6$. The levels which are underlined
           correspond to the 3-state Potts model (89) with 2 sites}
  \begin{tabular}{ccc} \hline
 \makebox[35 mm]{${\cal G}_{0}$}& \makebox[35 mm]{${\cal G}_{1}$}&
 \makebox[35 mm]{${\cal G}_{2}$} \\ \hline
 \underline{0.000000}& -                   & -              \\
    0.424413         & \underline{0.424413}& -              \\
    0.579759         & \underline{0.579759}& -              \\
 \underline{0.848826}& -                   & -              \\
    1.004172         & \underline{1.004172}& -              \\
    1.159518         & 1.159518            & \underline{1.159518}\\ \hline
  \end{tabular}
\end{table}

\subsection{Conclusion}

We have treated the closed quantum invariant chain for the quantum parameter
$q=e^{\imath\varphi}$, $\varphi=\frac{\pi m}{m+1}$, $m=3,4,\ldots$ . This
model has the conformal anomaly [7]
\bdm
c=1-\frac{6}{m(m+1)} \;\; .
\edm
The Hamiltonian also has the property that it implies sector-dependent
boundary conditions. We have shown that from the partition function of this
theory we can construct partition functions of well-defined statistical
systems. In particular, the spectra of these are subsets of the spectrum of the
quantum chain and can be obtained using (34). These formulae
have been obtained using the theory of representations of quantum groups and
keeping the ``good'' states and omitting the ``bad'' states.

We have shown how our construction is related to the well-known projection
mechanism of statistical models from Hamiltonians with toroidal boundary
conditions.

Finally, using this relation we have been able to obtain partition functions
in the finite-size scaling limit. This has enabled us to find the operator
content
of the systems constructed from the quantum chain. These systems belong to
the family of unitary minimal models. These properties have been illustrated
in a few particular cases ($m=3,4,5$).
\newpage
{\bf ACKNOWLEDGMENTS}\\
One of us (S.P.) would like to thank for permanent interest to
V.Rittenberg and for precious discussions to H.Grosse and P.Martin.

\newpage
\centerline{ {\bf REFERENCES}}
\bigskip
\begin{description}
\item{[1]}
  Pasquier V and Saleur H 1990 {\em Nucl.Phys.}B {\bf 330} 523
\item{[2]}
  Batchelor M T, Mezincescu L, Nepomechie R I and Rittenberg V 1990
      {\em J.Phys.A:} {\em Math.Gen.} {\bf 23} L141
\item{[3]}
  Meljanac M, Milekovi\c\ M and Pallua S 1991 {\em J.Phys.A:Math.Gen.}
      {\bf 24} 581
\item{[4]}
  Batchelor M T and Kumba K 1991 {\em J.Phys.A:Math.Gen.} {\bf 24} 2599
\item{[5]}
  Martin P and Rittenberg V 1992 {\em Int.J.Nucl.Phys.}A, Vol.7 Suppl 1B, 797
\item{[6]}
  Martin P P 1991 {\em Potts Models and Related Problems in Statistical
      Mechanics} (Singapore: World Scientific)
\item{[7]}
  Grosse H, Pallua S, Prester P and Raschhofer E 1994 {\em J.Phys.A:Math.Gen.}
      {\bf 27} 4761
\item{[8]}
  Karowski M and Zapletal A 1994 {\em Nucl.Phys.}B {\bf 419} [FS] 567
\item{[9]}
  Alcaraz F C, Barber M N and Batchelor M N 1988 {\em Ann.Phys.} {\bf 182} 280
\item{[10]}
  Alcaraz F C, Grimm U and Rittenberg V 1989 {\em Nucl.Phys.}B {\bf 316} 735
\item{[11]}
  Grimm U and Sch\"{u}tz G 1993 {\em J.Stat.Phys.} {\bf 71} 923
\item{[12]}
  Sch\"{u}tz G 1993 {\em J.Phys.A:Math.Gen.} {\bf 26} 4555
\item{[13]}
  J\"{u}ttner G and Karowski M 1994 {\em Completeness of ``Good'' Bethe Ansatz
\\
      Solutions of a Quantum Group Invariant Heisenberg Model} preprint
      hep-th/ \\ 9406183
\item{[14]}
  Reshetikhin N and Smirnov F 1990 {\em Commun.Math.Phys.} {\bf 131} 157
\item{[15]}
  Rocha-Caridi A 1985 {\em Vertex Operators in Mathematics and Physics} ed
      J \\ Lepowski, S Mandelstam and I Singer (Berlin: Springer)
\item{[16]}
 Capelli A, Itzykson C and Zuber J B 1987 {\em Nucl.Phys.}B {\bf 280} [FS18]
445
\item{[17]}
  von Gehlen G and Rittenberg V 1986 {\em J.Phys.A:Math.Gen.} {\bf 19} L625
\item{[18]}
  von Gehlen G, Rittenberg V  and Ruegg H 1985 {\em J.Phys.A:Math.Gen.}
      {\bf 19} 107
\end{description}

\end{document}